\documentclass[a4paper]{mem}
\usepackage{natbib}
\usepackage{graphicx}
\usepackage[a4paper]{hyperref}
\begin{document}
   \title{Distance scale and variable stars \\ in Local Group Galaxies: LMC and
   Fornax}

   \author{M. Maio \inst{1}, L. Baldacci \inst{1}, G. Clementini \inst{1}, 
          C. Greco \inst{1}, M. Gullieuszik \inst{2}, E.V. Held \inst{2}, 
          E. Poretti \inst{3}, L. Rizzi \inst{2},
          A. Bragaglia \inst{1}, E. Carretta \inst{2}, 
          L.~Di~Fabrizio \inst{4}, R.~Gratton \inst{2}, 
          E. Taribello \inst{1} }

   \offprints{M. Maio}
\mail{\\ marcella@saigon.bo.astro.it}

   \institute{INAF - Osserv. Astron. di Bologna,
via Ranzani 1, 40127 Bologna, Italy \\ 
              \and  INAF - Osserv. Astron. di Padova,~Vicolo~
	      dell'Osservatorio~5,~35122~Padova,~Italy\\
	      \and INAF - Osserv. Astron. di Brera, Via Bianchi 46, 23807 Merate, Italy\\
              \and INAF - Centro Galileo Galilei \& Telescopio Nazionale Galileo, PO Box
	      565, 38700 Santa Cruz de La Palma, Spain\\
             }

   \abstract{
We briefly review our photometric and spectroscopic study
of RR Lyrae variable stars in the bar of the Large Magellanic Cloud (LMC), 
that allowed us to reconcile 
the so-called {\em short} and {\em long} distance
moduli of the LMC on the value $\mu_{LMC}=18.51 \pm 0.085$ mag. 
Then we present preliminary results 
from the photometric study of a 33'$\times$34' area 
in the Fornax dwarf spheroidal galaxy containing  
the stellar clusters Fornax 3 (NGC 1049) and 6.
We identified about 1000 candidate variables in this field of 
Fornax, 
and report the first detection and measure of about 60 RR Lyrae variable 
stars in the globular cluster Fornax 3.
   }
   \authorrunning{M. Maio et al.}
   \titlerunning{Distance scale and variable stars in LMC and Fornax}
   \maketitle
%

\vspace{-0.5cm}
\section{Introduction}
   \begin{figure*}
   \centering
   \includegraphics[height=.32\textheight, width=.5\textwidth, bb=35 164 600 718,clip]{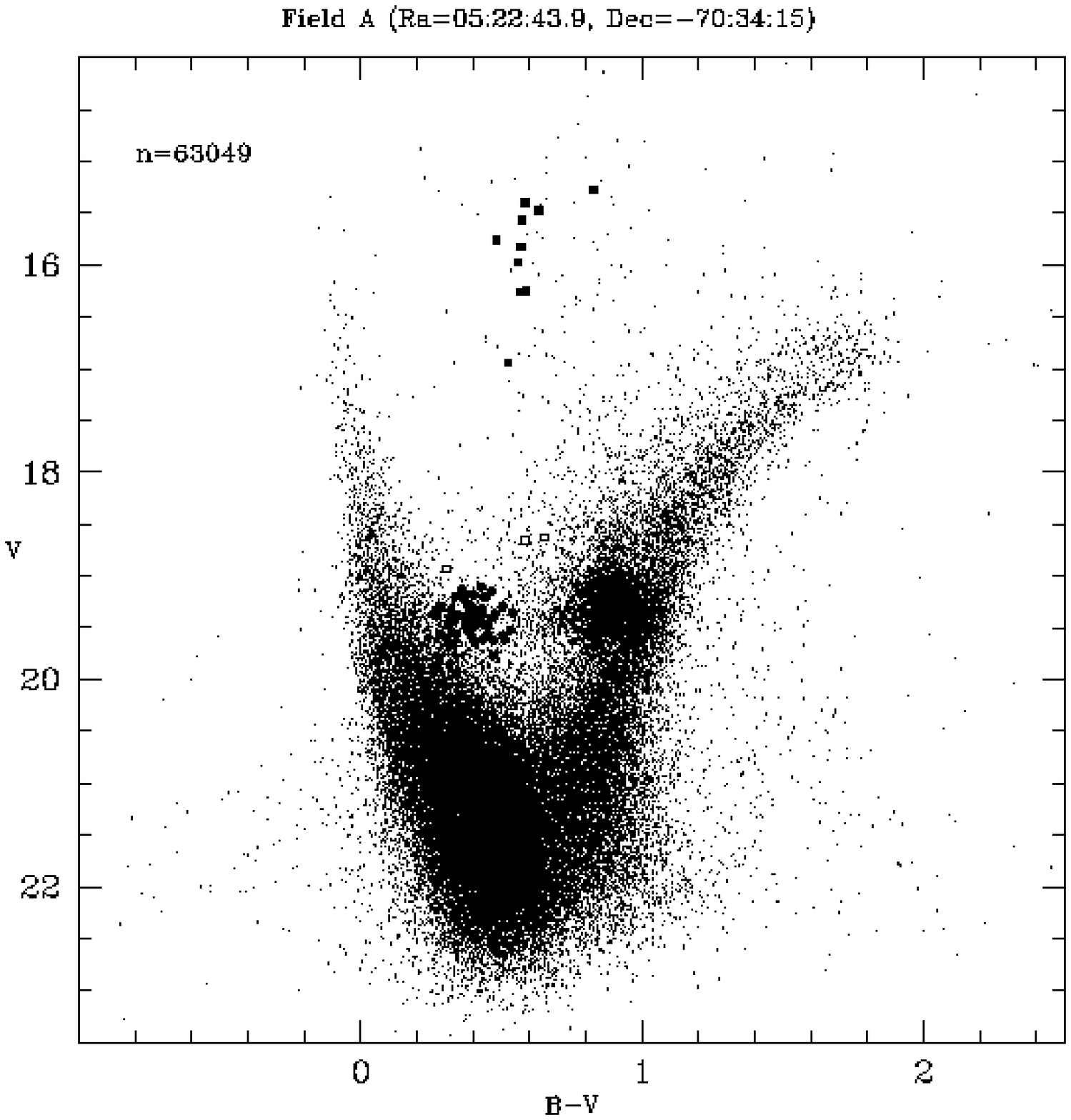}%
   \includegraphics[height=.32\textheight, width=.5\textwidth, bb=18 220 399 500,clip]{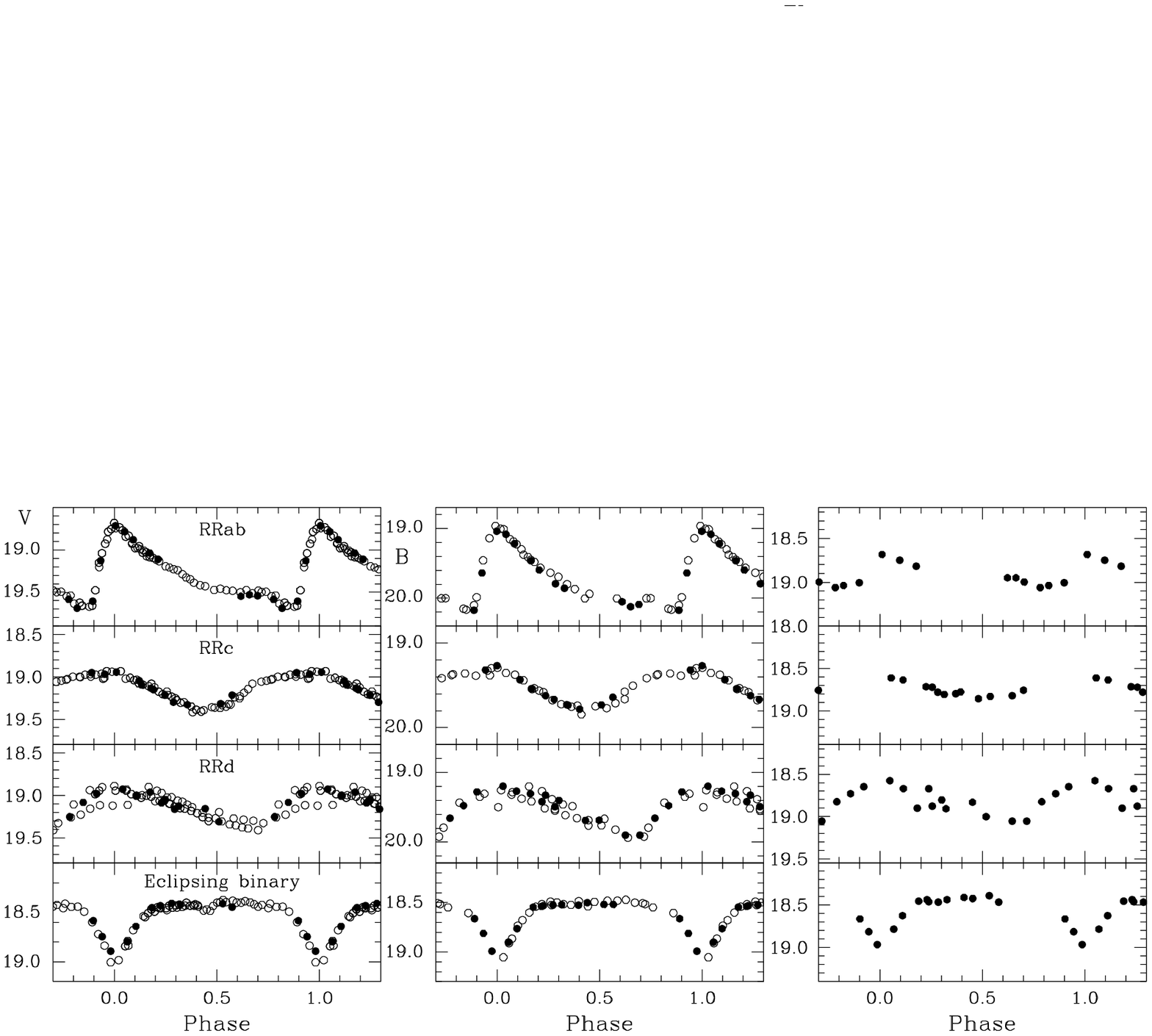}
   \caption{Left panel: V {\em vs} B $-$ V color - magnitude diagram of 
   our LMC Field A. Larger symbols mark the variables
   identified in this area. Right panel: examples of our light curves.}
              \label{fig1}
    \end{figure*}

Variable stars are important to set the astronomical distance scale, and 
to sample different stellar populations in galaxies.
RR Lyrae stars, in particular, trace the oldest stellar component and are the primary 
Population II distance indicators in the Local Group galaxies.

\vspace{-0.5cm}
\section{The Large Magellanic Cloud}

The distance to the LMC has for a long time remained uncertain
notwithstanding its vicinity to the Milky Way. 
In fact, the LMC distance modulus from different
indicators ranges from about 18.3 mag (e.g. Baade-Wesselink
and statistical
parallax methods: Fernley et al. 1998a,b) to about 18.7 mag (e.g.
Cepheid trigonometric parallaxes:
Feast \& Catchpole 1997).
In the following, we present the results of our study of variable stars in the LMC and 
its impact on the {\em short} and {\em long} distance scale controversy.

We obtained time series photometric data (72~$V$, 41~$B$, and 15~$I$ frames)
of two 13' $\times$
13' fields (Field A and B) close to the bar of the LMC 
and derived light curves accurate to 0.02-0.03 mag, for 
125 RR Lyrae stars (77 RRab's, 38 RRc's, 10 double-mode pulsators RRd's), 
4 anomalous Cepheids, 11 classical Cepheids, 11 eclipsing
binaries, and one $\delta$ Scuti star.

   \begin{figure*}
   \centering
   \resizebox{\textwidth}{!}
   {\includegraphics[height=.34\textheight, width=.41\textwidth, bb=38 150 575 750,clip]{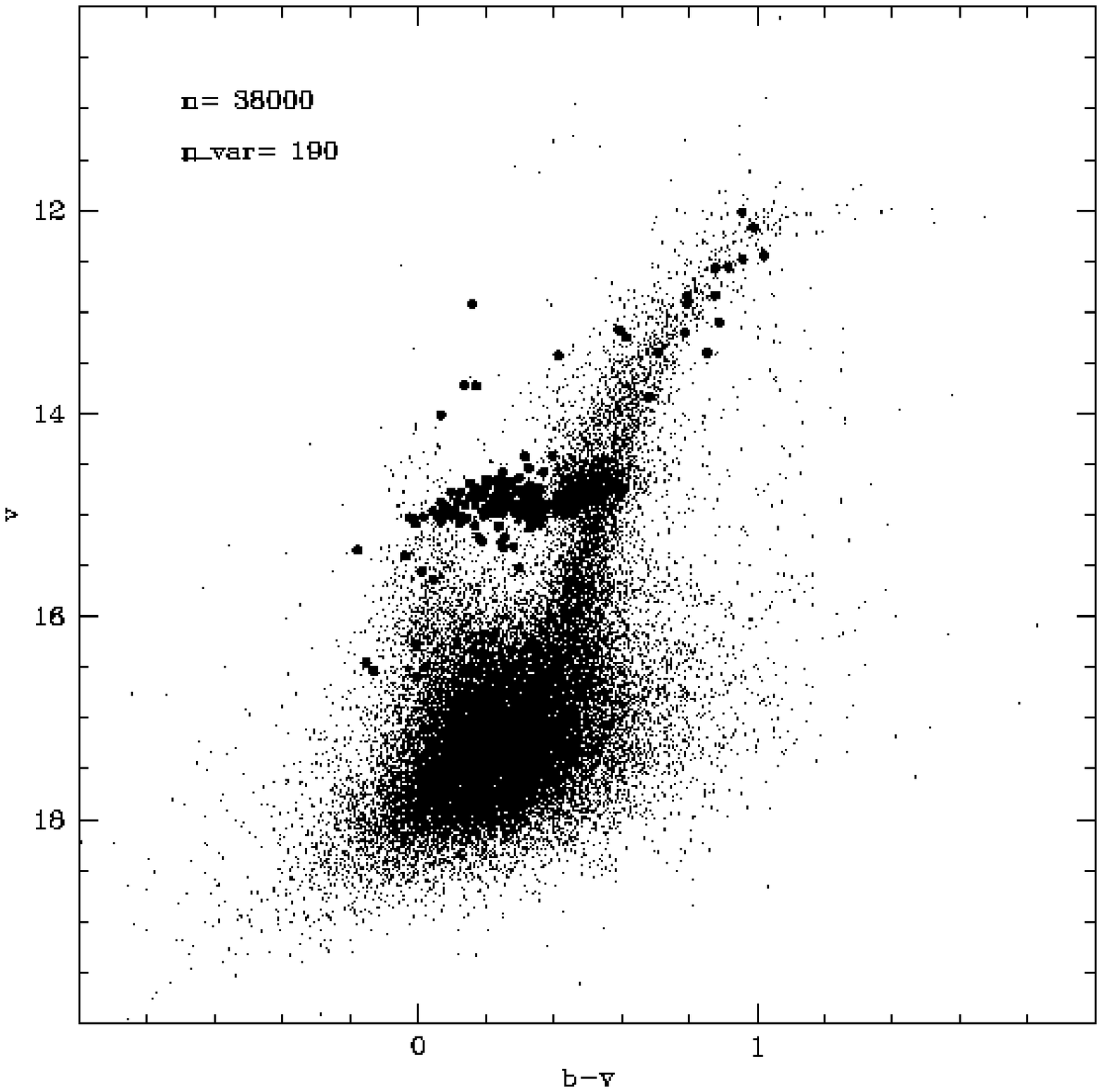}
   \includegraphics[height=.29\textheight, width=.4\textwidth, bb=142 222 469 530,clip]{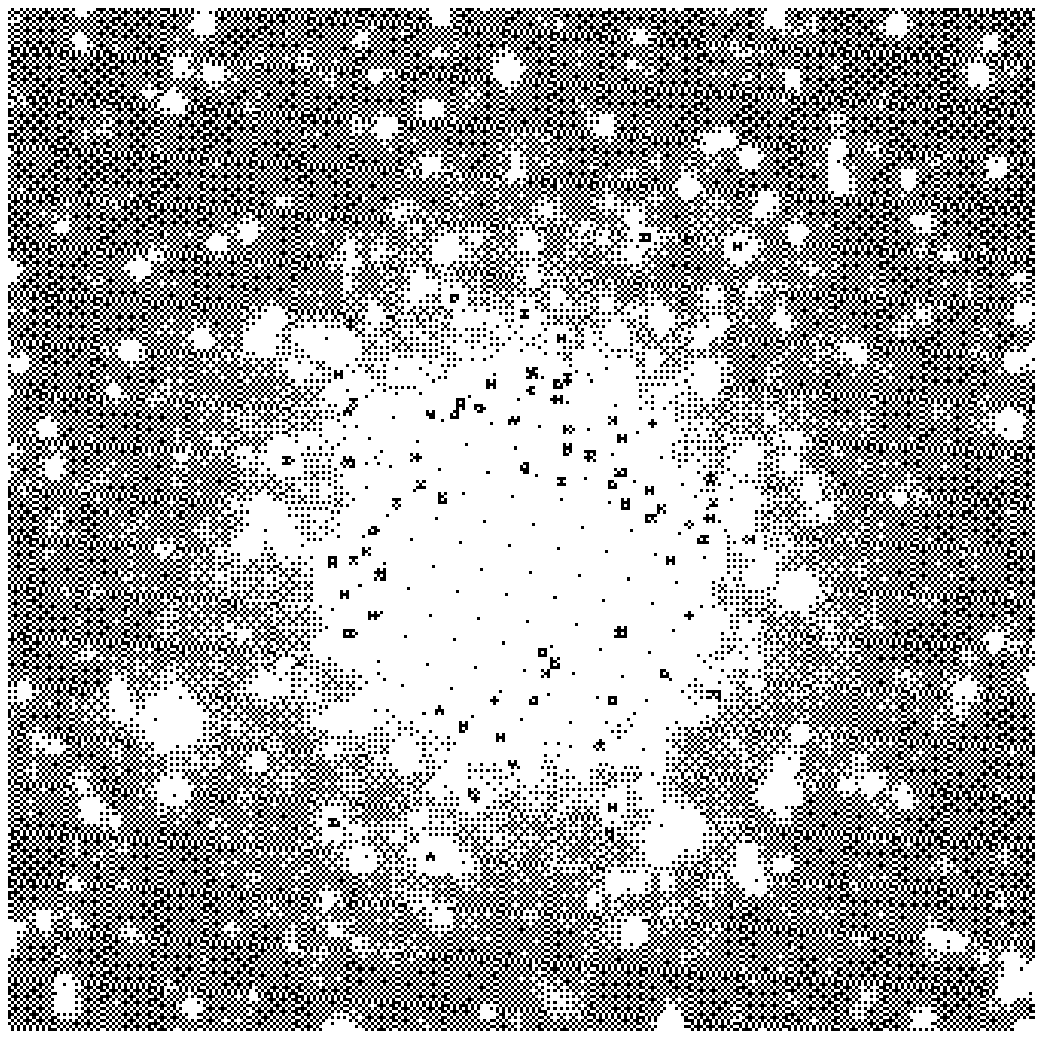}
   \includegraphics[height=.32\textheight, width=.35\textwidth, bb=19 220 215 510,clip]{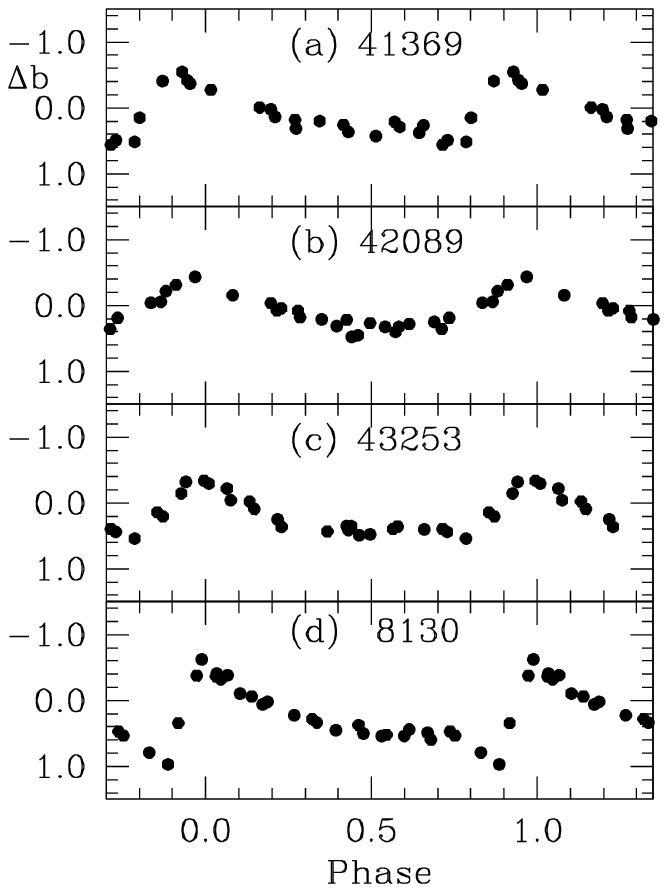}     
   }
   \caption{On the left: Instrumental color-magnitude diagram of chip \#7; candidate variables
   are marked as larger symbols.
   In the center: image 95" $\times$ 95" of the globular cluster NGC 1049, with candidate
   variables marked as filled squares.
   On the right: examples of light curves in Fornax 3, [panels (a) and (b)], and in the field,
   [panels (c) and (d)].
   }
              \label{fig2}
    \end{figure*}

Figure \ref{fig1} shows, on the left, the location of the variable stars in the HR
diagram of Field A; variables are plotted according to their intensity-average
magnitudes and colors. On the right, we show the light curves for an RRab, an RRc,
an RRd and an eclipsing binary in our sample.

We also derived metallicities for a total number of about 100 of the 
RR Lyrae stars using different methods. 
We applied the $\Delta$S method (Preston 1959) to 6 RRd's
(Bragaglia et al. 2001), and a revised version of this technique  
to spectra obtained with 
FORS at the VLT in 2001 for 101 RR Lyrae stars in our sample.
Finally, photometric metallicities were estimated from  parameters of the Fourier
decomposition of the $V$ light curves (Jurksic \& Kov\'acs 1996,  Kov\'acs 
\& Walker 2001) for 29 RRab's.

All these different estimates are in very good agreement to each other, once differences 
in the adopted metallicity scales are properly taken into account and 
give an average metal abundance of [Fe/H]=$-$1.5.

We used the VLT spectroscopic metallicity determinations and our
estimates of the average luminosities of the RR Lyrae stars and reddening,
to derive the luminosity-metallicity relation followed by the LMC
RR Lyrae stars:
$$<V_0(RR)> = [0.214 (\pm 0.047)] \times({\rm [Fe/H]} +$$
$$1.5) + 19.064(\pm 0.017)$$
(Clementini et al. 2003, Gratton et al. 2003, in preparation).

Our estimate of the dereddened average luminosity of the RR Lyrae stars 
in the LMC bar is $<V(RR)>_0=19.06$ mag (at ${\rm[Fe/H]}= -1.5$ dex). This value was
combined with a number of recent independent estimates for the 
absolute magnitude of the RR Lyrae stars (e.g.
Gratton et al. 2002, Cacciari et al. 2000) to estimate the LMC distance. 
The distance moduli so derived where compared with the most recent and
accurate LMC distance determinations from several other Population I and II
distance indicators, reaching  1 $\sigma$ convergence on a distance modulus
of $\mu_{LMC}=18.51 \pm 0.085$ mag, as fully described in Clementini et 
al. (2003).
\vspace{-0.25cm}
\section {The~Fornax~dwarf~spheroidal~galaxy}

Located about 140 Kpc from the Milky Way, Fornax is a dwarf spheroidal galaxy (dSph) 
dominated by an intermediate-age stellar population
(Stetson et al. 1998, Saviane et al. 2000).
There is also evidence for an old stellar component, since the galaxy contains
5 globular clusters and field population as old and
metal-poor as that in globular clusters (Saviane et al. 2000). 
The field of Fornax dSph has been investigated for
variability by Bersier \& Wood (2002) who surveyed
a half square degree covering the central region 
and found about
600 variables (among which 515 RR Lyrae). However, because of
the small telescope size and the mediocre seeing
conditions of their observations, RR Lyrae are close to the detection 
limit of their photometry, and their observations did not result
in high quality light curves.
GCs of Fornax dSph have never been adequately
surveyed for variability, in spite of clear indications, from their
HB morphologies, that they should indeed contain RR Lyrae variables
(Buonanno et al. 1998,1999). 

Fornax 3, one of the two clusters that lie in our Fornax field,
is quite metal-poor ([Fe/H=$-$1.96 $\pm$0.20, Buonanno et al. 1998) and
has a relatively well populated Horizontal Branch (HB) characterized by
a very extended HB blue tail. Buonanno et al. (1998) identified
66 candidate RR Lyrae stars in this cluster from their limited HST data. 

We observed an area of 34' $\times$ 33' North to the Fornax dSph center
with the Wild Field Imager (WFI)
of the 2.2 m ESO-MPI telescope, with the larger part of the galaxy in chip 
\#6 and \#7 of the WFI mosaic. 
We obtained time series B and V photometry (18 $V$ and 62 $B$ frames).
Photometric reductions using the packages DAOPHOT/ALLSTAR II (Stetson 1998)
and ALLFRAME (Stetson 1994) are in progress.
Candidate variables were identified using the package ISIS 2.1 (Alard 2000)
that works with the image subtraction method. We detected
335 candidate variable stars in chip \#6 and 190 candidate variables  in chip \#7. 
The globular
cluster Fornax 3 falls in chip \#6. We selected a box of 95" $\times$ 95"
centered in the core of Fornax 3 and in this area we find about 70 candidate variables. Even
if decontamination from the field variables has not been made yet, these candidate variables 
are very likely to belong to the cluster, and since most of them fall on the cluster HB, they 
are RR Lyrae stars. This is the first detection and measure of the variable stars in one
of Fornax dSph galaxy globular cluster system. Figure \ref{fig2} shows
in the left panel the (v, v-b) instrumental color-magnitude diagram of Fornax dSph field in chip 
\#6, and in the central panel an image of the globular cluster Fornax 3. In both panels the
candidate variables are marked by filled circles. Finally, the right panel shows some
preliminary light curves of RR Lyrae stars in Fornax GC3 (panels {\em a} and {\em b})
and Fornax dSph field (panels {\em c} and {\em d}). These are DAOPHOT instrumental differential
{\em b} light curves and only half of the time series data has been plotted. Photometric 
reductions are still in progress and we expect that the photometric quality of the RR Lyrae
light curves (about 0.05-0.06 mag for each data points in the DAOPHOT reductions) will
improve to a few hundredths of a magnitude for the magnitude calibrated Alard fluxes. We also
detected 4 candidate variables in the scarcely populated cluster Fornax 6, an object whose
actual nature still needs to be investigated.

We estimate that the total number of candidate variable stars in the 8 chips of our WFI
field of Fornax galaxy is of about 1000 (lower limit), to be compared to the 600 found 
by Bersier \& Wood 2002 in an area about 1.6
times larger than ours. This should be taken into when extrapolating Bersier \& Wood 
results to determine the total census of the variable star population in Fornax dSphs
galaxy.
\vspace{-0.5cm}
\bibliographystyle{aa}

\begin{thebibliography}{}

\bibitem[{Alard 2000}]{a00} Alard, C. 2000, A\&AS, 144, 363
\bibitem[{Bersier \& Wood 2002}]{bw02} Bersier, D. \& Wood, P.R. 2002, AJ, 123,
         840
\bibitem[{Bragaglia et al. 2001}]{b01} Bragaglia, A., Gratton, R.G., Carretta,
         E., Clementini, G., Di Fabrizio, L. \& Marconi, M. 2001, AJ, 122, 207
\bibitem[{Buonanno et al. 1998}]{b98} Buonanno, R., Corsi, C.E., Zinn, R., Fusi
         Pecci, F., Hardy, E. \& Suntzeff, N.B. 1998, AJ, 501, L33 
\bibitem[{Buonanno et al. 1999}]{b99} Buonanno, R., Corsi, C.E., Castellani, M.,
         Marconi, G., Fusi Pecci, F., Zinn, R. 1999, AJ, 118, 1671 
\bibitem[{Cacciari et al. 2000}]{c02} Cacciari, C., Clementini, G., Castelli,
         F. \& Melandri, F. 2000, in ASP Conf. Ser. 203, The Impact of Large
	 Scale Surveys on Pulsating Stars Research, ed. L. Szabados \& D. Kurtz
	 (San Francisco: ASP), ASP, 176
\bibitem[{Clementini et al. 2003}]{c03} Clementini, G., Gratton, R.G., Bragaglia,
         A., Carretta, E., Di Fabrizio, L. \& Maio, M. 2003, AJ, 125, 1309
\bibitem[{Fernley et al. 98}]{fa98} Fernley, J., Barnes, T. G., Skillen, I.,
          Hawley, S. L., Hanley, C. J., Evans, D. W., Solano, E. \& Garrido, R. 1998,
          A\&A, 330, 515 
\bibitem[{Fernley et al. 98}]{fa98} Fernley, J., Skillen, I., Carney, B. W.,
         Cacciari, C. \& Janes, K. 1998, MNRAS, 293, L61
\bibitem[{Feast et al. 97}]{fa97} Feast, M. W. \& Catchpole, R. M. 1997, MNRAS, 286, L1
\bibitem[{Gratton et al. 2002}]{g02} Gratton, R.G., Bragaglia, A., Carretta, E.,
         Clementini, G. \& Grundahl, F. 2002, in ASP Conf. Ser., New Horizons
	 in Globular Clusters Astronomy, G. Piotto, G. Meylan, G. Djorvski \&
	 M. Riello (San Francisco: ASP), in press
\bibitem[{Gratton et al. 2003}]{g03} Gratton, R.G., Bragaglia, A., Clementini,
         G., Carretta, E., Taribello, E., Di Fabrizio, L. \& Maio, M. 2003, AJ,
	 in preparation
\bibitem[{Jurksic \& Kov\'acs 1996}]{j96} Jurksic, J., \& Kov\'acs, G. 1996,
         A\&A, 312, 111
\bibitem[{Kov\'acs \& Walker 2001}]{k01} Kov\'acs, G., \& Walker, A.R., 2001,
         A\&A, 371, 579
\bibitem[{Preston 1959}]{p59} Preston, G.W., 1959, ApJ, 130, 507
\bibitem[{Saviane et al. 2000}]{s00} Saviane, I., Held, E.V. \& Bertelli, G.
         2000, A\&A, 355, 56. 
\bibitem[{Stetson 1994}]{s94} Stetson, P.B. 1994, PASP, 106, 250
\bibitem[{Stetson 1998}]{s98} Stetson, P.B. 1998,"User's Manual~for\\
         ~DAOPHOT~II"~\\http://www.star.bris.ac.uk/~mbt/daophot
\bibitem[{Stetson et al. 1998}]{setal98} Stetson, P.B., Hesser, James E.,
          Smecker-Hane, Tammy A. 1998, PASP, 110, 533
\end{thebibliography}

\end{document}